\documentclass[reprint,aps]{revtex4-1}

\usepackage{graphicx}
\usepackage{amsmath}
\usepackage{hyperref}
\usepackage{tabularx}
\usepackage{hyphenat}

\begin{document}
\title{Characterizing the attenuation of coaxial and rectangular microwave-frequency waveguides at cryogenic temperatures}
\author{P. Kurpiers$^1$, T. Walter$^1$, P. Magnard$^1$, Y. Salathe$^1$, A. Wallraff$^1$}
\affiliation{$^1$Department of Physics, ETH Z\"urich, CH-8093, Z\"urich, Switzerland.}
\date{\today}

\begin{abstract}
\nohyphens{Low-loss waveguides are required for quantum communication at distances beyond the chip-scale for any low-temperature solid-state implementation of quantum information processors. We measure and analyze the attenuation constant of commercially available  microwave-frequency waveguides down to millikelvin temperatures and single photon levels. More specifically, we characterize the frequency-dependent loss of a range of coaxial and rectangular microwave waveguides down to $0.005\,\rm{dB}/\rm{m}$ using a resonant-cavity technique. We study the loss tangent and relative permittivity of commonly used dielectric waveguide materials by measurements of the internal quality factors and their comparison with established loss models. The results of our characterization are relevant for accurately predicting the signal levels at the input of cryogenic devices, for reducing the loss in any detection chain, and for estimating the heat load induced by signal dissipation in cryogenic systems.}
\end{abstract}
\maketitle

\section{Introduction}
\label{sec:Intro}

Interconverting the quantum information stored in stationary qubits to photons and faithfully transmitting them are two basic requirements of any physical implementation of quantum computation~\cite{DiVincenzo2000}. Coherent interaction of solid-state and atomic quantum devices with microwave photons has been experimentally demonstrated for quantum dot systems~\cite{Petta2005,Frey2012,Maune2012, Delbecq2011}, individual electron spin qubits~\cite{Pla2012}, ensembles of electronic spins~\cite{Kubo2010}, superconducting circuits~\cite{Chiorescu2004,Wallraff2004,Devoret2013} and Rydberg atoms~\cite{Hagley1997,Raimond2001,Haroche2006,Hogan2012}.

In the field of circuit quantum electrodynamics, experiments show the ability to use single itinerant microwave photons~\cite{Eichler2011,Eichler2012b} or joint measurements~\cite{Roch2014} to generate entanglement between distant superconducting qubits~\cite{Narla2016}. In these probabilistic entanglement schemes the entanglement generation rate is inversely proportional to the signal loss between the two sites. Furthermore, entanglement can be generated deterministically by transmitting single microwave photons with symmetric temporal shape~\cite{Cirac1997} which can be emitted~\cite{Pechal2014,Zeytinoglu2015} and reabsorbed with high fidelity~\cite{Wenner2014}. However,  the fidelity of the entangled state is dependent on the signal loss for most protocols. Therefore, the ability to transmit microwave photons with low loss, which we address in this manuscript, is essential for the realization of quantum computation with solid-state and atomic quantum systems.

In addition, studying the reduction of loss of superconducting waveguides has the potential to contribute to improving the fidelity of qubit state measurements~\cite{DiVincenzo2000} by minimizing the loss of the signal between the read-out circuit and the first amplifier~\cite{Macklin2015}. Knowing the loss of microwave waveguides also enables more accurate estimates of the signal levels at the input of cryogenic devices and could be used to better  evaluate the heat load induced by signal dissipation.

Previous studies of the attenuation constant were performed for different types of superconducting coaxial cables down to $4\,\rm{K}$ by  impedance matched measurements~\cite{McCaa1969, Ekstrom1971, Chiba1973, Giordano1975, Mazuer1978, Peterson1989, Kushino2008, Kushino2013}. In those works, the attenuation constant is typically evaluated from measurements of the transmission spectrum of the waveguide, which is subsequently corrected for the attenuation in the interconnecting cables from room temperature to the cold stage in a reference measurement. In these studies lengths of the low-loss superconducting waveguides between $20\,\rm{m}$ and $400\,\rm{m}$ were used for the measurements to be dominated by the device under test.

In this paper we study the loss of coaxial cables and rectangular waveguides using a resonant-cavity technique from which we extract attenuation constants down to $0.005\,\rm{dB}/\rm{m}$ accurately between room and cryogenic temperatures at the tens of millikelvin level. By utilizing higher-order modes of these resonators we measure the frequency dependence of the attenuation for a frequency range between $3.5$ and $12.8\,\rm{GHz}$ at cryogenic temperatures only limited by the bandwidth of our detection chain.
By comparing our data to loss models capturing this frequency range we extract the loss tangent and relative permittivity of the dielectric and an effective parameter characterizing the conductor loss.

We evaluate the attenuation constant of coaxial and rectangular waveguides made by a number of different manufacturers from a range of materials, see Table~\ref{tab:overviewCoaxWaveguides}. We characterize $2.2\,\rm{mm}$ ($0.085\,\rm{in}$) diameter coaxial cables with Niobium-Titanium (Nb-$47$ weight percent $(\rm{wt}\%)$ Ti) or Niobium outer and center conductors. For both cables the manufacturer Keycom Corporation~\cite{Keycom2016} used a low density polytetrafluorethylen (ldPTFE) dielectric. We also measure an aluminum outer, silver plated copper wire (SPC) \footnote{per the standard specification for silver-coated soft or annealed copper wire (ASTM B-298)} center conductor coaxial cable with an ldPTFE dielectric and an outer diameter of $3.6\,\rm{mm}$ ($0.141\,\rm{in}$) manufactured by Micro-Coax, Inc.~\cite{MicroCoax2016}. As a reference, we analyze a standard copper outer, silver plated copper clad steel (SPCW) \footnote{per the standard specification for silver-coated, copper-clad steel wire (ASTM B-501)} center conductor coaxial cable with a solid PTFE (sPTFE) dielectric and an outer conductor diameter of $2.2\,\rm{mm}$ ($0.085\,\rm{in}$) manufactured by Micro-Coax, Inc.~\cite{MicroCoax2016}. We investigate rectangular waveguides of type WR90 by Electronic Industries Alliance (EIA) standard with inner dimensions of $s_{1}=22.86\,\rm{mm}$, $s_{2}=10.16\,\rm{mm}$ ($s_{1}=0.900\,\rm{in}$, $s_{2}= 0.400\,\rm{in}$) with a recommended frequency band of $8.2$ to $12.4\,\rm{GHz}$. Three different conductor materials are characterized: aluminum 6061 with chromate conversion coating per MIL-C-5541E, aluminum 6061 without further surface treatment and oxygen-free, high conductivity (OFHC) copper with tin (Sn $\rm{wt}\%>99.99\%$) plating of thickness $5-10\,\mu\rm{m}$ on the inner surface. All three rectangular waveguides are manufactured by Penn Engineering Components, Inc.~\cite{Pennengineering2016}.

\begingroup
\squeezetable
\begin{table*}[t]
\begin{tabularx}{\textwidth}{ X | X | X | X | X | X | X | X}
\hline\hline
ID  		           & CC085NbTi      & CC085Nb        & CC141Al       & CC085Cu        & WR90Alc        & WR90Al     & WR90CuSn \\
\hline
dim. [$\rm{mm}$ ($\rm{in}$)]  &$2.2$ ($0.085$)&$2.2$ ($0.085$)&$3.6$ ($0.141$)&$2.2$ ($0.085$) & WR90      & WR90   & WR90   \\
conductor                          & NbTi/NbTi     & Nb/Nb         & Al/SPC	       & Cu/SPCW        & coated Al & Al     & Cu-Sn  \\
dielectric                         & ldPTFE		     & ldPTFE		     & ldPTFE		     & sPTFE 		      & vacuum    & vacuum & vacuum \\
length [$\rm{mm}$ ($\rm{in}$)]     & $110$         & $110$         & $900$   & $120$& $304.8$ ($12$)&$304.8$ ($12$)&$304.8$ ($12$)\\
$T$(BT) [$\rm{mK}$]	& $120$        & $50$          & $60$          & $15$          & $60$           & $25$      & $50$          \\
$T$(4K) [$\rm{K}$]  & $4.0$        & $4.0$         & $4.0$         & $4.1$         & $4.3$          & $4.0$      & $4.0$        \\
$\nu$ range  [$\rm{GHz}$]& $4.2$-$12.7$&$4.1$-$12.5$& $3.5$-$12.7$&$3.6$-$12.7$&$7.9$-$12.3$&$7.7$-$12.8$&  $7.7$-$12.8$      \\
$\overline{n}$(BT)                 & 0.1-2         & 0.3-10        & 1-2           & 1-3            & 0.2-1     & 1-4    & 1-3   \\
$\overline{n}$(4K)                 & 0.4-5         & 0.2-4         & 8-16          & 4-9            & 2-10      & 3-12   & 5-20   \\
\hline\hline
\end{tabularx}
\caption{Summary of waveguide and measurement parameters. The indicated dimension (dim.) specifies the outer diameter of the coaxial cables and the EIA type of the rectangular waveguides. The conductor and dielectric materials are specified as well as the length of the resonant section employed for the measurements. The temperature $T$ measured at the waveguide is indicated.  The average photon number on resonance is shown for the investigated frequency ($\nu$) range.}
\label{tab:overviewCoaxWaveguides}
\end{table*}
\endgroup

\section{Experimental setup}
\label{sec:ExpSetup}

\begin{figure}[b]
\centering
\includegraphics[width=0.42\textwidth]{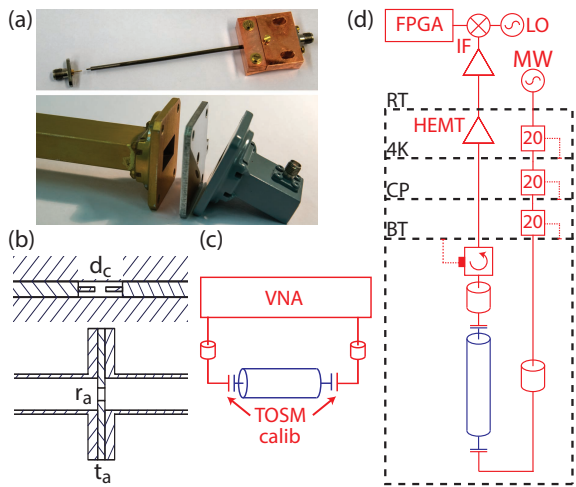}
\caption{Photographs (a) and schematics (b) of a capacitively coupled coaxial cable and an aperture coupled rectangular waveguide. (c) Schematic of the room temperature setup using a vector network analyzer (VNA). A through-open-short-match (TOSM) calibration is used to account for loss and phase offsets in the interconnecting cables. (d) Schematic of the FPGA-based microwave setup \cite{Lang2014} used for measurements at cryogenic temperatures.}
\label{fig:photoSetupRed}
\end{figure}

We construct resonators from coaxial cables and rectangular waveguides as shown in the photographs and schematics of Fig.~\ref{fig:photoSetupRed} (a) and (b). For the coaxial cables we use sub-miniature version A (SMA) panel mount connectors and remove the outer conductor and dielectric material of the coaxial cable at both ends to realize a capacitive coupling between the center conductor of the cable and the connector. We choose a coupling capacitance to obtain largely undercoupled resonators (see Sec.~\ref{sec:Coax} and Appendix~\ref{app:Coupling} for details).

At room temperature (RT) we use a vector network analyzer (VNA) and a through-open-short-match (TOSM) calibration to set the measurement reference plane to the input of the coupling ports of the waveguide according to the schematic presented in Fig.~\ref{fig:photoSetupRed} (c) and adjust the input and output coupling to be approximately equal. For measurements at cryogenic temperatures the microwave signal propagates through a chain of attenuators of $20\,\rm{dB}$ each at the 4 K, the cold plate and the base temperature stages before entering the waveguide (Fig.~\ref{fig:photoSetupRed} (d)). The output signal is routed through an isolator with a frequency range of $4$-$12\,\rm{GHz}$ and an isolation larger than $20\,\rm{dB}$, a high-electron-mobility transistor (HEMT) amplifier with a bandwidth of $1$-$12\,\rm{GHz}$, a gain of $40\,\rm{dB}$ and a noise temperature of $5\,\rm{K}$, as specified by the manufacturer.  After room temperature amplification and demodulation, the signal is digitized and the amplitude is averaged using a field programmable gate array (FPGA) with a custom firmware.

The waveguides are characterized at a nominal temperature of $4\,\rm{K}$ (4K) using the pulse tube cooler of a cryogen-free dilution refrigerator system in which also the millikelvin temperature (BT) measurements are performed. We thermally anchor the waveguides to the base plate of the cryostat using OFHC copper braids and clamps. The actual waveguide temperatures are extracted in a measurement of the resistance of a calibrated  Ruthenium oxide (RuO) sensor mounted at the center of the coaxial cables or at the end of the rectangular waveguides and are listed in Table~\ref{tab:overviewCoaxWaveguides}.

For the measurements at base temperature BT ($\sim 10\,\rm{mK}$) it proved essential to carefully anchor all superconducting waveguide elements at multiple points to assure best possible thermalization. The measured temperatures listed in Table~\ref{tab:overviewCoaxWaveguides} are found to be significantly higher than the BT specified above. We attribute the incomplete thermalization of the superconducting waveguides  to the small thermal conductivity of the employed materials below their critical temperature $T_{\rm{c}}$~\cite{Pobell2006}. We note that when using only a minimal set of anchoring points, we observed even higher temperatures.

\section{Measurements of the attenuation constant}
\label{sec:Coax}

\subsection{Illustration of the measurement technique}
\label{subsec:MeasTech}

\begin{figure}[b]
\centering
\includegraphics[width=0.48\textwidth]{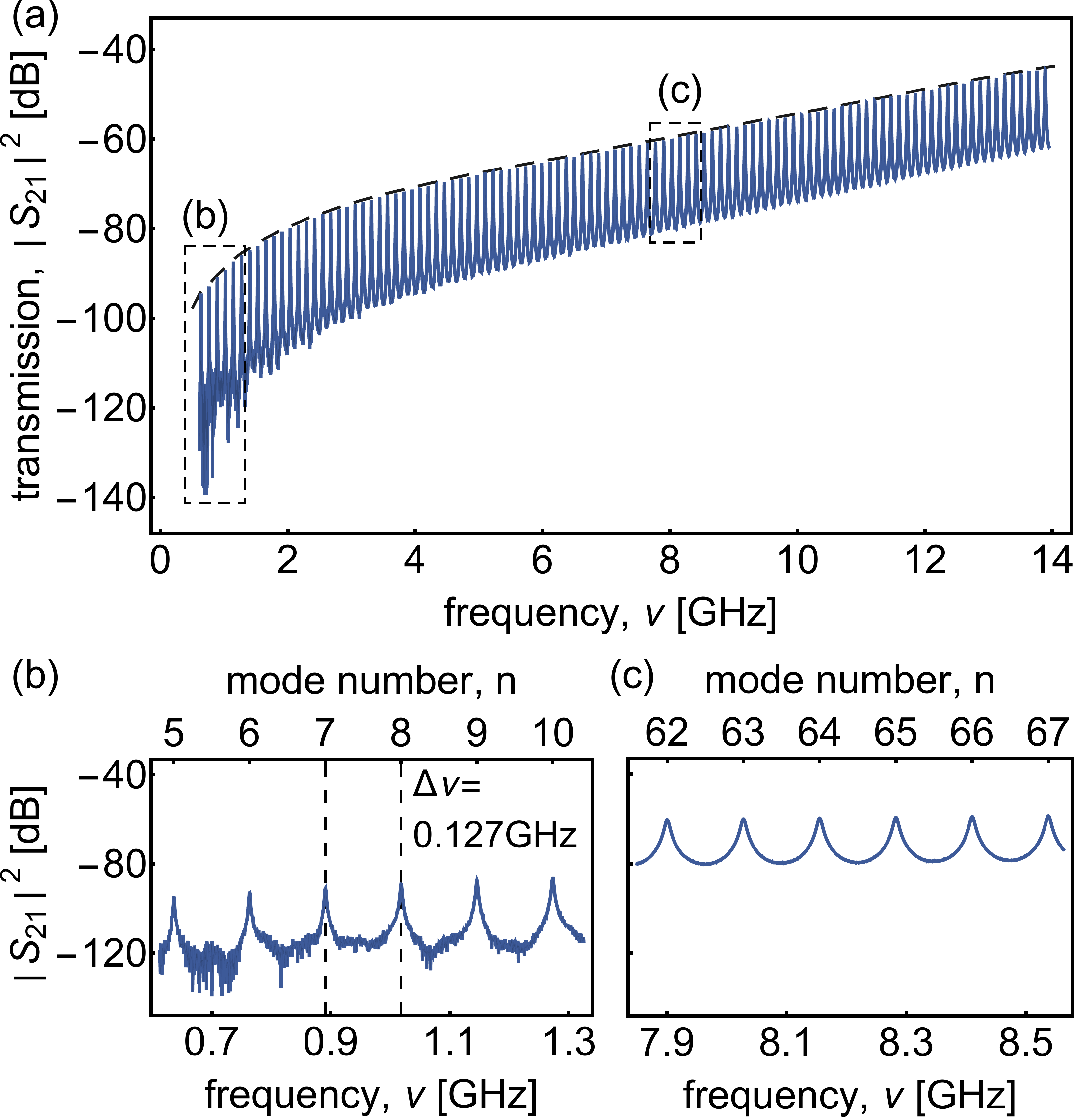}
\caption{(a) Transmission coefficient $|S_{21}|^2$ versus frequency $\nu$ for CC141Al at RT. The dashed line indicates the frequency-dependent insertion loss on resonance $IL(\nu_{\rm{n}})$. $|S_{21}|^2$ of (b) the first $6$ measured modes and (c) $6$ modes around $8\,\rm{GHz}$ as indicated by the dashed boxes in (a).}
\label{fig:plotSpectrum}
\end{figure}

\begin{figure}[b]
\centering
\includegraphics[width=0.48\textwidth]{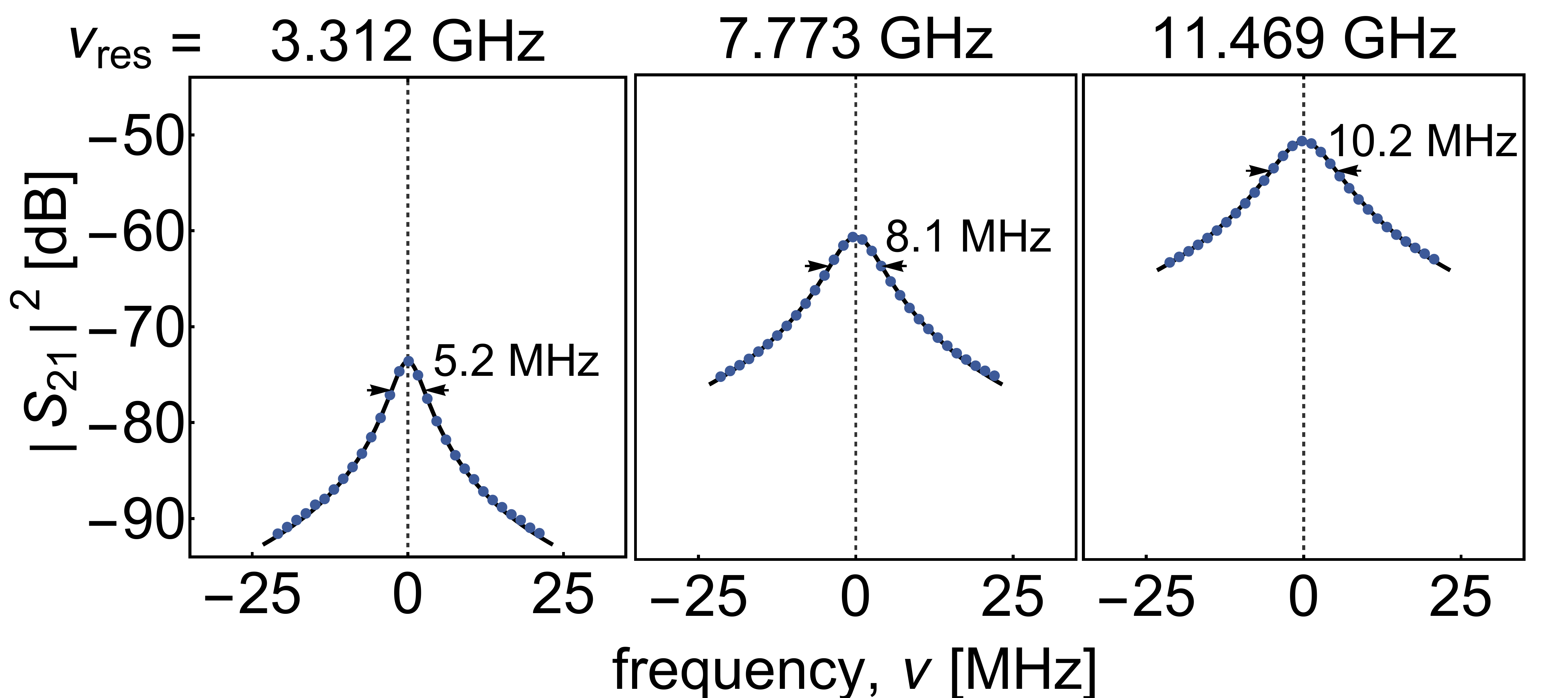}
\caption{Absolute value squared of the measured transmission $|S_{21}|^2$ (dots) versus frequency $\nu$ at the indicated resonances  for CC141Al (RT) and modes numbers $n=\{26,\;61,\;90\}$. The arrows indicate the full width at half maximum from which we extracted $Q_{\rm{i}}$. $\nu_{\rm{res}}$ is the center frequency of the resonance. The line is the absolute value squared of the simultaneous fit of the real and imaginary part of the $S_{21}$ scattering parameter (see Appendix~\ref{app:RTfit} for details).}
\label{fig:plotAbsRT}
\end{figure}

To illustrate the resonant-cavity technique for extracting the attenuation constant of a waveguide we discuss a calibrated S-parameter measurement at RT for the coaxial line CC141Al (Table~\ref{tab:overviewCoaxWaveguides}). The measured transmission spectrum $|S_{21}(\nu)|^2$ exemplifies the periodic structure of higher-order modes for mode numbers $n$ between $5$ and $109$ (Fig.~\ref{fig:plotSpectrum}). We extract the resonance frequency $\nu_{\rm{n}}$ and the external and internal quality factor, $Q_{\rm{e}}$ and $Q_{\rm{i}}$, for each mode $n$ by fitting the complex transmission coefficient of a weakly coupled parallel RLC circuit (see Appendix~\ref{app:RTfit}) to the data in a finite bandwidth around each $\nu_{\rm{n}}$ (Fig.~\ref{fig:plotAbsRT}). We observe a decreasing insertion loss on resonance $IL(\nu_{\rm{n}})=-10 \log_{10} |S_{21}(\nu_{\rm{n}})|^2 \, \rm{dB} $ (dashed line in Fig.~\ref{fig:plotSpectrum} (a)) with increasing frequency due to the increase of the effective capacitive coupling strength. We chose $IL(\nu_{\rm{n}}) > 40 \, \rm{dB}$ to ensure the largely undercoupled regime ($Q_{\rm{e}} \gg Q_{\rm{i}}$) over the entire frequency range. In this regime, $Q_{\rm{i}}$ is well approximated by the loaded quality factor $Q_{\rm{l}}$ according to $1/Q_{\rm{i}}=1/Q_{\rm{l}}-1/Q_{\rm{e}}\approx 1/Q_{\rm{l}}$. In our experiments we assure that $Q_{\rm{e}} > 10 \, Q_{\rm{i}}$ for all frequencies and temperatures.

Under this condition it is sufficient to extract $Q_{\rm{l}}$ for each mode $n$ from
\begin{equation}
|S_{21}(\nu)|=\frac{S^{\rm{max}}_{\rm{n}} }{\sqrt{1+4(\nu/\nu_{\rm{n}}-1)^2 {Q_{\rm{l}}}^2}}+C_{1}+C_{2} \nu
\label{eqn:S21abs}
\end{equation}
neglecting the specific value of the insertion loss ($S_{\rm{n}}^{\rm{max}}$ is a free scaling factor). $C_{1}$, $C_{2}$ account for a constant offset and a linear frequency dependence in the background~\cite{Petersan1998} most relevant for measurements of low quality factors ($<10^3$) resonances. 

\begin{figure}[b]
\centering
\includegraphics[width=0.48\textwidth]{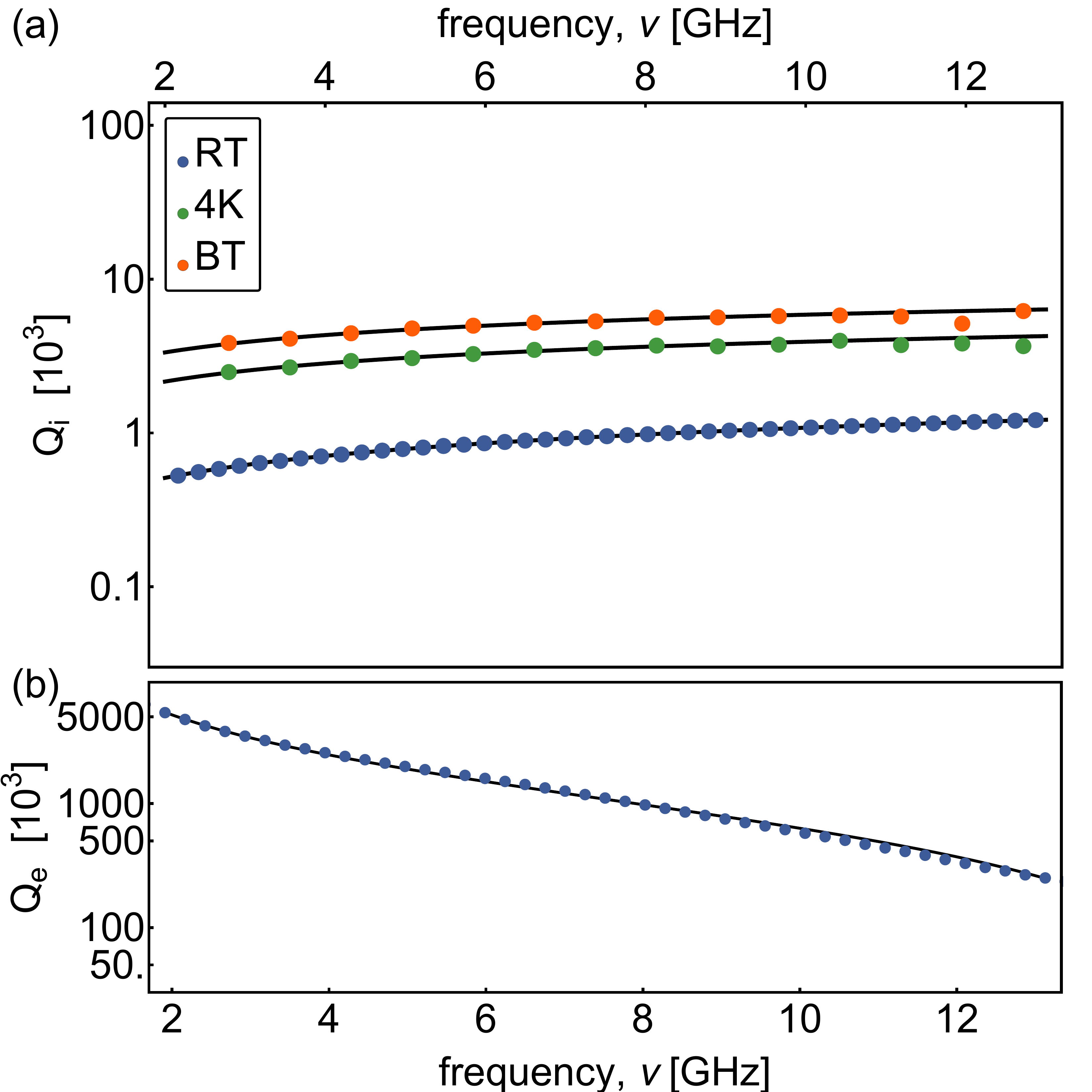}
\caption{(a) Measured internal ($Q_{\rm{i}}$) and (b) external quality factors ($Q_{\rm{e}}$) of coaxial line CC141Al versus frequency $\nu$ extracted from the spectra measured at RT (blue dots), at 4K (green dots) and BT (orange dots). The black lines are fits to the loss model discussed in the text.}
\label{fig:plotQintQext141AlSPC}
\end{figure}

\subsection{Analysis of coaxial lines}
\label{subsec:CoaxLines}

To determine the frequency dependence of the attenuation constant $\alpha(\nu)$ of the coaxial line we analyze its measured quality factors $Q_{\rm{i}}$ in dependence on the mode number $n$. The fundamental frequency $\nu_{0}$ of a low-loss transmission line resonator is given by
\begin{equation}
\nu_{0}= \frac{c}{\sqrt{\epsilon_{\rm{r}}}}\frac{1}{2 l}
\label{eqn:nu0}
\end{equation}
with the length of the resonator $l$, the relative permittivity of the dielectric $\epsilon_{\rm{r}}$ and the speed of light in vacuum $c$. The internal quality factor $Q_{\rm{i}}$~\cite{Pozar2012}
\begin{equation}
Q_{\rm{i}}=\frac{n \pi}{2 l \alpha}
\label{eqn:Qint}
\end{equation}
is inversely proportional to $\alpha$
\begin{equation}
\alpha=\alpha_{\rm{c}}+\alpha_{\rm{d}}
=\frac{g_{\rm{c}}\sqrt{\epsilon_{\rm{r}}}}{2 \mu_0 c} R_{\rm{s}}(\nu)+ \frac{\pi \sqrt{\epsilon_{\rm{r}}}}{c} \nu \, \rm{tan} \: \delta \, .
\label{eqn:alpha}
\end{equation}
$\alpha$ can be written as a sum of conductor loss $\alpha_{\rm{c}}$ and dielectric loss $\alpha_{\rm{d}}$ with the vacuum permeability $\mu_0$, a frequency dependent surface resistance $R_{\rm{s}}(\nu)$, a geometric constant $g_{\rm{c}}$ and the  frequency independent loss tangent of the dielectric material $\rm{tan} \: \delta$. For a coaxial line $g_{\rm{c}}$ is $(1/a+1/b)/\rm{ln}(b/a)$ with the radius of the center conductor $a$ and the inner radius of the outer conductor $b$. To characterize the conductor loss of coaxial cables combining different materials for the center and outer conductors we introduce an effective surface resistance $R_{\rm{s}}$ (see Appendix~\ref{app:LossCoax}). Inserting Eq.~\eqref{eqn:alpha} into Eq.~\eqref{eqn:Qint} leads to
\begin{equation}
Q_{\rm{i}}(\nu_{\rm{n}})=\frac{1}{\frac{g_{\rm{c}}}{2 \pi \mu_0}\frac{R_{\rm{s}}(\nu_{\rm{n}})}{\nu_{\rm{n}}}+\rm{tan} \: \delta}
\label{eqn:Qint2}
\end{equation}
which is independent of $\epsilon_{\rm{r}}$. Therefore, $\epsilon_{\rm{r}}$ is extracted from the fundamental frequency of the resonator $\nu_0$ and $R_{\rm{s}}(\nu)$ and $\rm{tan} \: \delta$ from measurements of $Q_{\rm{i}}(\nu)$.

The surface resistance of a normal conductor $R_{\rm{s}}^{\rm{nc}}(\nu)$ is proportional to $\sqrt{\nu}$ and to the direct current (dc) conductivity $1/\sqrt{\sigma}$~\cite{Pozar2012}. The theory of the high-frequency dissipation in superconductors~\cite{Tinkham2004, Mattis1958,Kose1989,Gao2008b} shows a quadratic dependence of $R_{\rm{s}}^{\rm{sc}}(\nu)$.

\begin{figure}[b]
\centering
\includegraphics[width=0.48\textwidth]{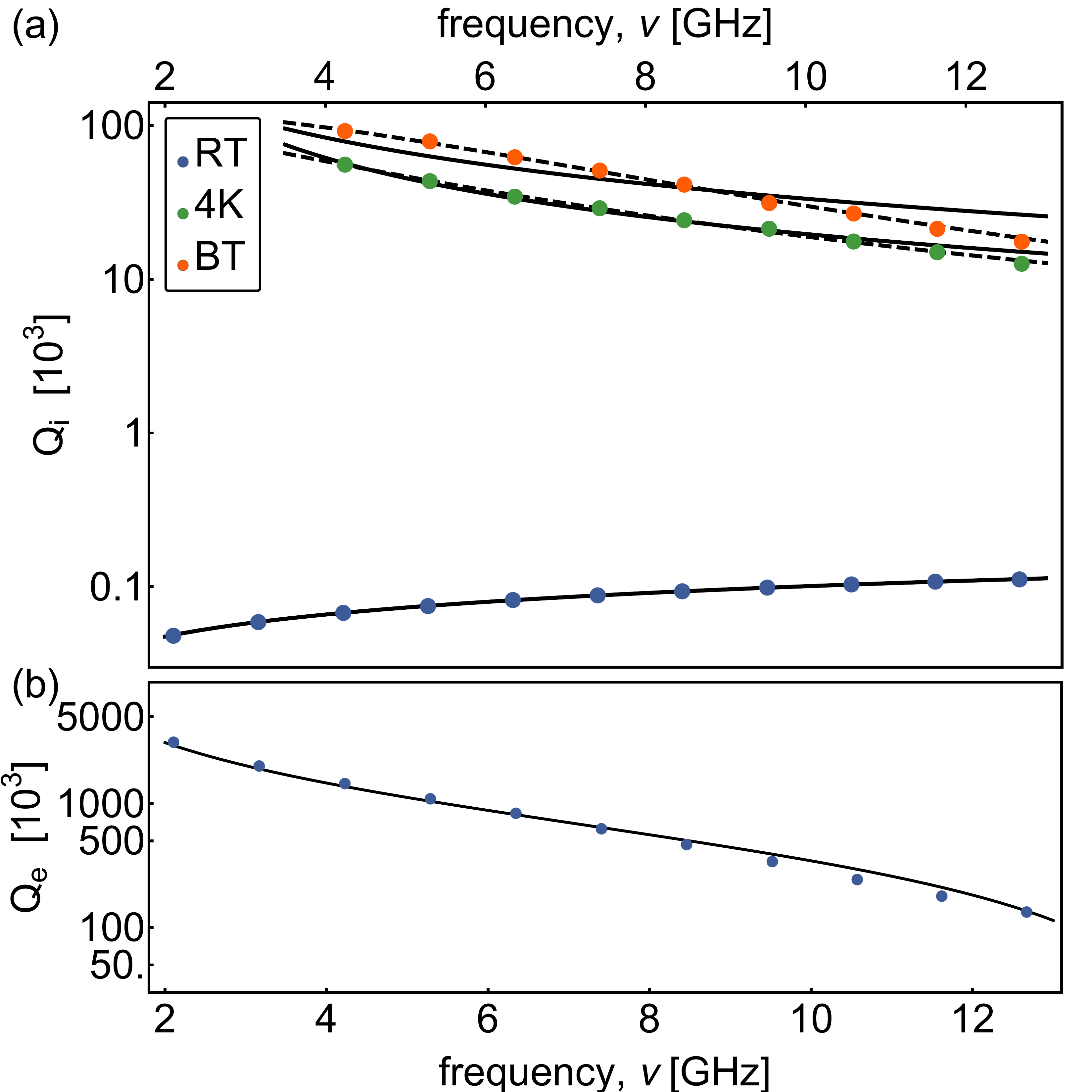}
\caption{(a) Measured ($Q_{\rm{i}}$) internal and (b) external quality factors ($Q_{\rm{e}}$) of coaxial line CC085NbTi versus frequency $\nu$ extracted from the spectra at RT (blue dots), at 4K (green dots) and BT (orange dots). The solid (dashed) black lines are fits to the loss model which assumes $R_{\rm{s}}^{\rm{sc}}(\nu) \propto \nu^2$ $(R_{\rm{s}}^{\rm{sc}}(\nu) \propto \nu^{\rm{p}})$.}
\label{fig:plotQintQext85NbTiNbTi}
\end{figure}

The measured external quality factors at RT (Fig.~\ref{fig:plotQintQext141AlSPC} (b) and Fig.~\ref{fig:plotQintQext85NbTiNbTi} (b)) are in good agreement with the ones expected for a capacitively coupled transmission line \cite{Goppl2008}
\begin{equation}
Q_{\rm{e}}(\nu_{\rm{n}})=\frac{C_{\rm{l}} l}{8 \pi C_{\rm{c}}^2 R_{\rm{l}}}\frac{1}{\nu_{\rm{n}}}+\frac{R_{\rm{l}} C_{\rm{l}} l \pi}{2} \nu_{\rm{n}}\,,
\label{eqn:Qext}
\end{equation}
with the capacitance per unit length $C_{\rm{l}}$, the real part of the load impedance $R_{\rm{l}}$ and the coupling capacitance $C_{\rm{c}}$ used as fit parameters. An interpolation of the $Q_{\rm{e}}$ measurements is used in the 4K and BT measurements to estimate the average number of photons stored in the waveguide on resonance at each mode $n$ (Table~\ref{tab:overviewCoaxWaveguides}).

The frequency dependence of the measured quality factors for CC141Al presented in Fig.~\ref{fig:plotQintQext141AlSPC} (a) shows the expected $\sqrt{\nu}$ dependence considering an effective conductivity of the outer and center conductor following the skin effect model of normal conductor. This suggests that $\alpha_{\rm{c}}$ is mainly limited by the normal conducting SPC center conductor. The dielectric loss limit of $Q_{\rm{i}}$ is determined to be approximately $15 \times 10^{3}$ at BT, see Table~\ref{tab:overviewEpsilonR}.

\begin{figure}[b]
\centering
\includegraphics[width=0.48\textwidth]{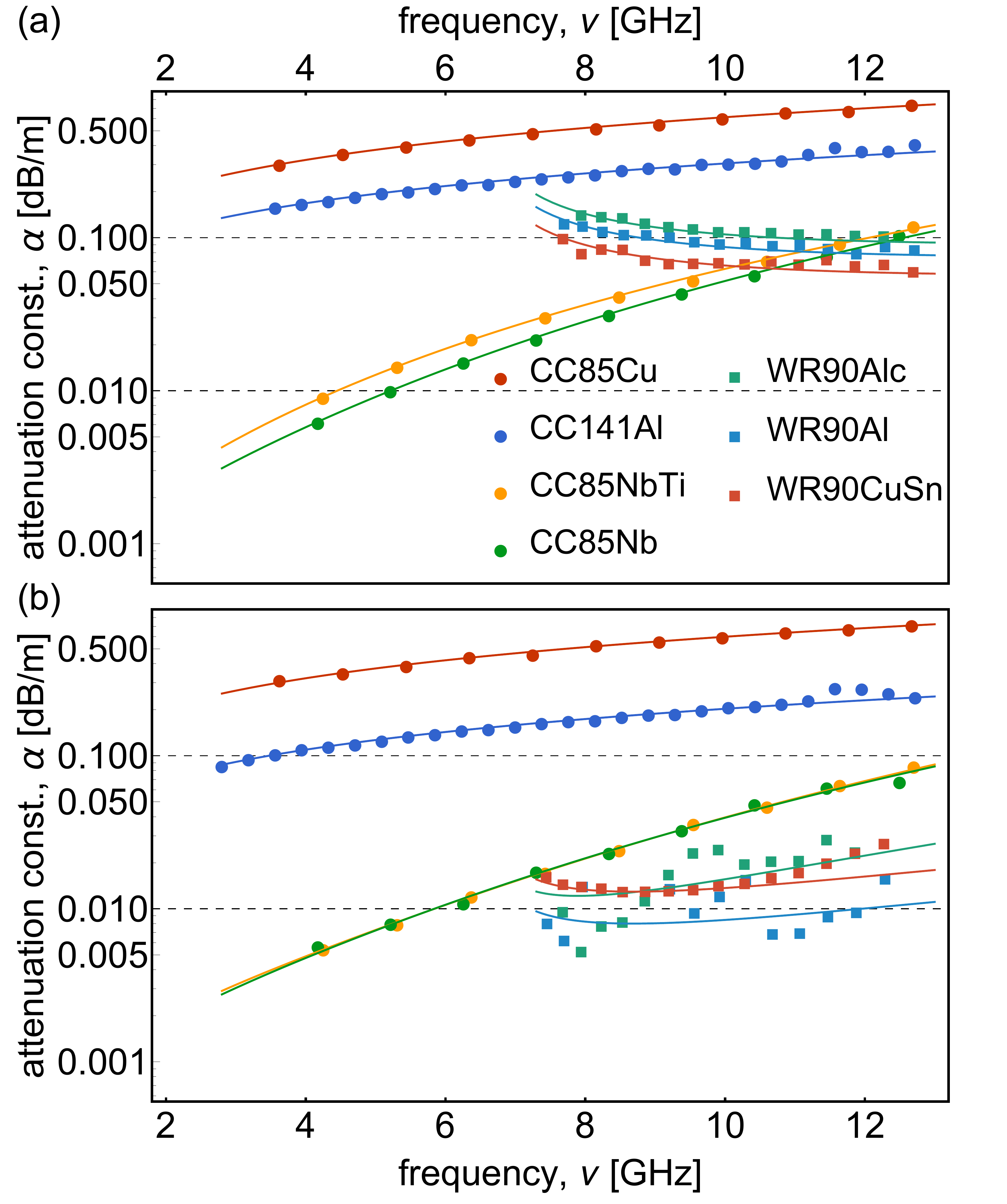}
\caption{Attenuation constant $\alpha$ (dots) extracted from measurements at frequency $\nu$ for CC085Cu, CC141Al, CC085NbTi, CC085Nb, WR90Alc, WR90Al and WR90CuSn. (a) shows the measurement results at approximately $4\,\rm{K}$ and (b) the measurement results at BT. The attenuation constant $\alpha$ is plotted in $\rm{dB}/\rm{m}$ in a log-plot to cover the full range of measured losses. The solid lines are calculated from the fits to $Q_{\rm{i}}$ measurements using the same model as for the data.}
\label{fig:OverviewAlpha}
\end{figure}

Following the same measurement procedure, we extract the quality factor of low-loss superconducting cables (e.g. see Fig.~\ref{fig:plotQintQext85NbTiNbTi} for CC085NbTi). The measured internal quality factors of CC085NbTi at the 4K and BT, ranging from $12\times10^3$ to $92\times10^3$, decrease approximately $\propto \nu^2$ (solid line) with a small deviation at higher frequencies. We obtain a better fit assuming a power law dependence of $R_{\rm{s}}^{\rm{sc}}(\nu) \propto \nu^{\rm{p}}$ with an exponent $p \approx 2.7 \pm 0.3$ at 4K and $p \approx 3.4 \pm 0.5$ at BT. This peculiar frequency dependence is not explained by the theory of high-frequency dissipation in superconductors~\cite{Tinkham2004, Mattis1958,Kose1989,Gao2008b}. Measuring CC085Nb leads to similar results as for CC085NbTi (Fig.~\ref{fig:OverviewAlpha}). We also observe a power law dependence with  $p \approx 3.2 \pm 0.4$ and $p \approx 3.3\pm 0.3$ at 4K and BT. Furthermore, we compare the dielectric and conductor properties of these low-loss coaxial cables with those of CC085Cu  for which me measured attenuation ranging from $0.30 \, \rm{dB}/\rm{m}$ to $0.75 \, \rm{dB}/\rm{m}$ (Fig.~\ref{fig:OverviewAlpha}). In addition, we measured the attenuation constant of a stainless steel outer and center conductor coaxial cable (CC085SS) at RT, approximately $77\,\rm{K}$ (LN2) and $4.2\,\rm{K}$ (LHe) described in Appendix~\ref{app:CC085SS}.

\begingroup
\squeezetable
\begin{table}[t]
\begin{tabularx}{0.48\textwidth}{X  X X X}
		\hline\hline
			T/parameter			&		Micro-Coax \newline LD PTFE 	&	 Keycom \newline ldPTFE  			&	Micro-Coax \newline sPTFE					\\\hline
			\multicolumn{4}{l}{$\epsilon_{\rm{r}}$}	\\	\hline
			RT 							& 	$1.70\pm0.01$								&	$1.72\pm0.06$								& $1.98\pm0.07$								\\
			4K 							& 	$1.70\pm0.01$								&	$1.72\pm0.06$								& $2.01\pm0.07$								\\
			BT 							& 	$1.70\pm0.01$								&	$1.72\pm0.06$								& $2.01\pm0.07$								\\\hline
			\multicolumn{4}{l}{$\rm{tan} \: \delta \ [\times 10^{-5}]$}	\\	\hline			
			RT 							& 	$9\pm1 $										&															& $25\pm4$										\\
			4K 							& 	$8.5\pm0.2$									&	$0.8\pm0.2$									& $22\pm2$										\\
			BT 							& 	$6.6\pm0.2$									&	$0.7\pm0.1$									& $19\pm4$										\\
			\hline\hline
\end{tabularx}
\caption{Overview of the extracted relative permittivities $\epsilon_{\rm{r}}$ and loss tangents $\rm{tan} \: \delta$ of the tested dielectric materials. The methods used for the extraction of these parameters are discussed in the text.}
\label{tab:overviewEpsilonR}
\end{table}
\endgroup

We extract the relative permittivities $\epsilon_{\rm{r}}$ from Eq.~\ref{eqn:nu0} and the loss tangent $\rm{tan} \: \delta$ from fitting Eq.~\ref{eqn:Qint2} to the measured $Q_{\rm{i}}(\nu)$ for each coaxial cable (Table~\ref{tab:overviewEpsilonR}). The values for the Micro-Coax ldPTFE are determined from CC141Al measurements, for the Micro-Coax sPTFE from CC085Cu and for the Keycom ldPTFE from CC085NbTi and CC085Nb measurements. We extract $\rm{tan} \: \delta$ of the Keycom ldPTFE from the fit assuming $R_{\rm{s}}^{\rm{sc}}(\nu) \propto \nu^{\rm{p}}$ at 4K and BT. Due to the low internal quality factors $Q_{\rm{i}}<100$ limited by the low RT conductivity (measured $\sigma\approx 8\times10^{5} \, \rm{S}/\rm{m}$ for NbTi and Nb) we are unable to extract these quantities at RT. At cryogenic temperature we observe that $\rm{tan} \: \delta$ of the ldPTFE of Micro-Coax and Keycom differ by a factor of $\sim10$. $\epsilon_{\rm{r}}$ is found to be $\sim1.7$ for ldPTFE and $\sim2$ for sPTFE and is nearly temperature independent.

\subsection{Analysis of rectangular waveguides}
\label{subsec:RecWaveguides}

We performed similar measurements with three different rectangular waveguides of type WR90 (see Table~\ref{tab:overviewCoaxWaveguides}). We use an aperture coupling approach by installing two Aluminum 1100  plates (thickness $t_{\rm{a}}=3\,\rm{mm}$) at both ends with a circular aperture (radius $r_{\rm{a}}=5.3\,\rm{mm}$ for WR90Alc and $r_{\rm{a}}=4.1\,\rm{mm}$ for WR90Al and WR90CuSn) in the center (Fig.~\ref{fig:photoSetupRed}) resulting in inductively coupled rectangular 3D cavities~\cite{Collin1991}. The coupling strength depends on $r_{\rm{a}}$ and $t_{\rm{a}}$ of the aperture plates. We perform finite element simulation to estimate the coupling (for details see Appendix~\ref{app:Coupling}) and determine the attenuation constant of the rectangular waveguides by a measurement of its internal quality factor.
For rectangular waveguide cavities the frequencies of the transverse electric modes $\rm{TE}_{10\rm{k}}$ are given by
\begin{equation}
\nu_{\rm{k}}=\frac{c}{2}\sqrt{\frac{1}{s_1^2}+\frac{k^2}{l^2}}
\label{eqn:fTe10k}
\end{equation}
with length of the longer transverse dimension of the rectangular waveguide $s_{1}$ and the length of the cavity $l$. The frequency dependent internal quality factor is~\cite{Pozar2012}
\begin{equation}
\begin{split}
&Q_{\rm{TE}_{10\rm{k}}}(\nu_{\rm{k}})=\\
&\frac{2 s_1^3 s_2 l \pi \mu_0 \nu_{\rm{k}}^3}{R_{\rm{s}}(\nu_{\rm{k}})}\frac{1}{c^2 s_2 (l-s_1)+2 s_1^3(2 s_2+l) \nu_{\rm{k}}^2 }
\label{eqn:Qint3Dcavity}
\end{split}
\end{equation}
with the length of the shorter transverse dimension of the rectangular waveguide $s_{2}$. Inverting Eq.~\eqref{eqn:Qint3Dcavity} we extract the surface resistance $R_{\rm{s}}(\nu_{\rm{k}})$ from a measurement of $Q_{\rm{TE}_{10\rm{k}}}(\nu_{\rm{k}})$ which we use to calculate the attenuation constant of the $\rm{TE}_{10}$ mode of a rectangular waveguide
\begin{equation}
\alpha_{\rm{TE}_{10}}(\nu)=
\frac{R_{\rm{s}}(\nu)}{s_1^2 s_2 \mu_0 c}\frac{s_2 c^2+2 s_1^3 \nu^2}{\nu \sqrt{4\nu^2 s_1^2-c^2}}\,.
\label{eqn:alphaWaveguide}
\end{equation}
Using this model we extract the attenuation constant of the  rectangular waveguides, ranging from $0.06 \, \rm{dB}/\rm{m}$ to $0.17\, \rm{dB}/\rm{m}$ at 4K and $0.007 \, \rm{dB}/\rm{m}$ to $0.02\, \rm{dB}/\rm{m}$ at BT, and determine the frequency dependence of the internal quality factor. For the rectangular waveguides in the normal state at 4K we find good agreement to the theoretical model by considering the normal state surface resistance $R_{\rm{s}}^{\rm{nc}}(\nu) \propto \sqrt{\nu} $. At BT in the superconducting state a surface resistance $R_{\rm{s}}^{\rm{sc}}(\nu) \propto \nu^2$ approximates the data (Fig.~\ref{fig:OverviewAlpha}). Note that, the  frequency dependence cannot be extracted with high accuracy for superconducting rectangular waveguides, since $\alpha_{\rm{TE}_{10}}(\nu)$ diverges towards the cutoff frequency $\nu_{\rm{co}}=c/2s_1$.

\section{Conclusions}
\label{sec:Conclusion}

We have presented measurements of the attenuation constant of commonly used, commercially available low-loss coaxial cables and rectangular waveguides down to millikelvin temperatures in a frequency range between $3.5$ and $12.8\,\rm{GHz}$. We have performed measurements of attenuations constants down to $0.005\,\rm{dB}/\rm{m}$ using a resonant-cavity technique at cryogenic temperatures. In this method, we employ weak couplings to the waveguides resulting in resonant standing waves and measure their quality factors. We have extracted the loss tangent and relative permittivity of different dielectric materials by comparing our measurement results to existing loss models. The frequency dependence of the internal quality factors of the normal conducting waveguides are well described by the loss model, while the tested CC085NbTi and CC085Nb show small deviations from the predictions for the high-frequency dissipation in superconductors~\cite{Mattis1958,Tinkham2004}. We have also studied the power dependence of the attenuation constant which we find to be independent of the input power in a range from $-140$ to $-80\, \rm{dBm}$ (see Appendix~\ref{app:PowerDep}). 

Our results indicate that transmitting signals on a single photon level is feasible within laboratory distances, e.g.~$95\,\%$ of the signal can be transmitted over distances of $28\, \rm{m}$ using commercial rectangular waveguides or $8\, \rm{m}$ using coaxial cables. Furthermore, we find no significant dependence of the attenuation constants on the ambient residual magnetic fields in measurements performed with and without cryoperm magnetic shielding (see Appendix~\ref{app:MagFieldDep}).

Comparing our results to recent measurements of high quality 3D cavities~\cite{Reagor2013} with quality factors up to $7\times\, 10^7$ indicate that improving the surface treatment of rectangular waveguides may lead to a even lower attenuation constant of rectangular waveguides down to $\sim 10^{-4}\,\rm{dB}/\rm{m}$. Furthermore, our measurements show that the loss tangent $\rm{tan} \: \delta$ strongly dependents on the PTFE composite where $\rm{tan} \: \delta\sim  2\times 10^{-6}$ of PTFE have been reported at cryogenic temperatures \cite{Geyer1995, Jacob2002} about a factor of 4 lower than those measured here. This suggests that the loss of superconducting coaxial cables may also be further reduced.

\acknowledgments

The authors thank Tobias Frey, Silvia Ruffieux and Maud Barthélemy for their contributions to the measurements, Oscar Akerlund for his support with the numerical integration and Christopher Eichler for discussing the manuscript. This work is supported by the European Research Council (ERC) through the "Superconducting Quantum Networks" (SuperQuNet) project, by National Centre of Competence in Research "Quantum Science and Technology" (NCCR QSIT), a research instrument of the Swiss National Science Foundation (SNSF), by the Office of the Director of National Intelligence (ODNI), Intelligence Advanced Research Projects Activity (IARPA), via the U.S. Army Research Office grant W911NF-16-1-0071 and by ETH Zurich. The views and conclusions contained herein are those of the authors and should not be interpreted as necessarily representing the official policies or endorsements, either expressed or implied, of the ODNI, IARPA, or the U.S. Government. The U.S. Government is authorized to reproduce and distribute reprints for Governmental purposes notwithstanding any copyright annotation thereon.

\appendix

\section{Characterization of the input/output coupling}
\label{app:Coupling}

\begin{figure}[b]
\centering
\includegraphics[width=0.48\textwidth]{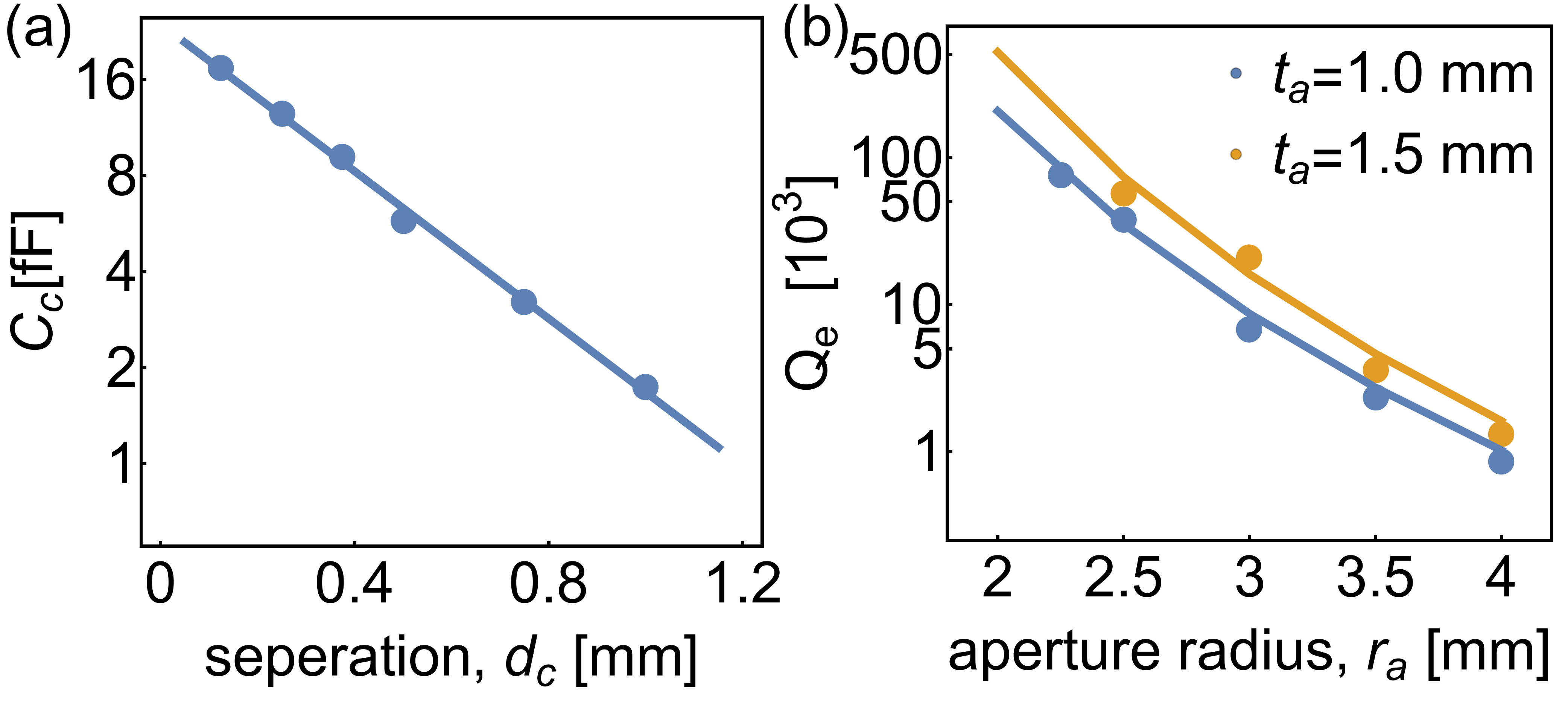}
\caption{a) Coupling capacitance $C_{\rm{c}}$ (dots) obtained from an ABCD transmission matrix model fit to the measured spectrum versus  the distance $d_{\rm{c}}$ between the center conductors of CC085Cu. The solid line is an exponential fit to the data. b) Measured $Q_{\rm{e}}$ of the $\rm{TE}_{101}$ mode of a 3D cavity aperture coupled to a rectangular waveguide. The solid line is obtained from a finite element simulation.}
\label{fig:plotCouplingAll}
\end{figure}

We extract the coupling capacitance $C_{\rm{c}}$ between the input coupler and the center conductor of the coaxial line CC085Cu from a fit of the measured transmission spectrum to an ABCD transmission matrix model. We find $C_{\rm{c}}$ to decrease exponentially with the separation $d_{\rm{c}}$, see Fig.~\ref{fig:photoSetupRed} and solid line in Fig. \ref{fig:plotCouplingAll} (a).

We determine the external quality factor $Q_{\rm{e}}$ of an aperture coupled 3D cavity with resonance frequency  $\nu_{101}=8.690\,\rm{GHz}$ in dependence on the coupling wall thickness $t_{\rm{a}}$ and the radius of the circular aperture $r_{\rm{a}}$ (Fig.~\ref{fig:photoSetupRed}). We find good agreement between the measured $Q_{\rm{e}}$ and the one extracted from finite-element simulations~\cite{COMSOL2012}, Fig.~\ref{fig:plotCouplingAll} (b).

By changing the geometry of the described couplers, the coupling of both coaxial cables and rectangular waveguides can be tuned  by orders of magnitude which is sufficient to fulfill the condition $Q_{\rm{e}}\gg Q_{\rm{i}}$ to extract $Q_{\rm{i}}$ precisely.

\section{Weakly coupled parallel RLC circuit}
\label{app:RTfit}

To extract the resonance frequency $\nu_{\rm{n}}$ and the external and internal quality factor, $Q_{\rm{e}}$ and $Q_{\rm{i}}$, of the investigated devices we simultaneously fit the real and imaginary part of the complex transmission coefficient
\begin{equation}
S_{21}(\nu)=(\frac{1}{1+Q_{\rm{e}}/Q_{\rm{i}}+2 i Q_{\rm{e}}(\nu/\nu_{\rm{n}}-1)}+X) e^{i \phi} \,.
\label{eqn:S21}
\end{equation}
of a weakly coupled parallel RLC circuit~\cite{Petersan1998}. Here $X$ is a complex constant which accounts for impedance mismatches in the SMA panel mount connectors and $e^{i \phi}$ is a rotation of the data relative to the measurement plane~\cite{Leong2002,Probst2015a}.

\section{Attenuation constant of low-loss coaxial cables}
\label{app:LossCoax}
We state the derivation of the attenuation constant of a low-loss transmission line based on Ref.~\cite{Pozar2012}. The equivalent circuit parameters (self-inductance per unit length $L_{\rm{l}}$,capacitance per unit length $C_{\rm{l}}$, series resistance per unit length $R_{\rm{l}}$ and shunt conductance per unit length $G_{\rm{l}}$) can be derived from the electric and magnetic field of the transmission line
\begin{eqnarray}
L_{\rm{l}}=\frac{\mu}{2\pi}\ln b/a, \nonumber \\
C_{\rm{l}}=\frac{2 \pi \epsilon_{\rm{0}} \epsilon_{\rm{r}}}{\ln b/a}, \nonumber\\
R_{\rm{l}}=\frac{1}{2\pi}(\frac{R^{\rm{a}}_{\rm{s}}}{a}+\frac{R^{\rm{b}}_{\rm{s}}}{b}), \nonumber\\
G_{\rm{l}}=\frac{2\pi \omega \epsilon_{\rm{0}} \epsilon_{\rm{r}} \rm{tan} \: \delta } {\ln b/a}.
\label{eqn:LCRG}
\end{eqnarray}
The complex propagation constant
\begin{equation}
\gamma =\alpha + i \beta=\sqrt{(R_{\rm{l}}+ i \omega L_{\rm{l}})(G_{\rm{l}}+i \omega C_{\rm{l}})}
\label{eqn:gamma1}
\end{equation}
can be approximated for small conductor $R_{\rm{l}} \ll \omega L_{\rm{l}}$ and dielectric loss $G_{\rm{l}} \ll \omega C_{\rm{l}}$ by
\begin{equation}
\gamma \approx i \omega \sqrt{L_{\rm{l}}C_{\rm{l}}} (1-\frac{i}{2}(\frac{R_{\rm{l}}}{Z_{\rm{0}}}+G_{\rm{l}} Z_{\rm{0}}))
\label{eqn:gamma2}
\end{equation}
with the characteristic impedance of the line $Z_{\rm{0}}=\sqrt{L_{\rm{l}}/C_{\rm{l}}}$, so that 
\begin{eqnarray}
\beta =\omega \sqrt{L_{\rm{l}}C_{\rm{l}}}=\frac{\omega \sqrt{\epsilon_{\rm{r}}}}{c}, \\
\alpha=\frac{1}{2}(\frac{R_{\rm{l}}}{Z_{\rm{0}}}+G_{\rm{l}} Z_{\rm{0}})=\nonumber\\
\frac{\sqrt{\epsilon_{\rm{r}}}}{2 c \ln b/a}(\frac{R^{\rm{a}}_{\rm{s}}}{a}+\frac{R^{\rm{b}}_{\rm{s}}}{b})+\frac{\pi \nu \sqrt{\epsilon_{\rm{r}}}}{c} \rm{tan} \: \delta=\nonumber\\
\frac{R_{\rm{s}}\sqrt{\epsilon_{\rm{r}}}}{2 c \ln b/a}(\frac{1}{a}+\frac{1}{b})+\frac{\pi \nu \sqrt{\epsilon_{\rm{r}}}}{c} \rm{tan} \: \delta.
\label{eqn:betaalpha}
\end{eqnarray}
In the last step we introduced an effective surface resistance $R_{\rm{s}}=R^{\rm{a}}_{\rm{s}}=R^{\rm{b}}_{\rm{s}}$ to characterize the conductor loss. This is necessary, since we cannot distinguish the contributions of the different materials of the center and outer conductors to the total conductor loss in our data.

\section{Attenuation constant of CC085SS}
\label{app:CC085SS}

\begin{figure}[t]
\centering
\includegraphics[width=0.48\textwidth]{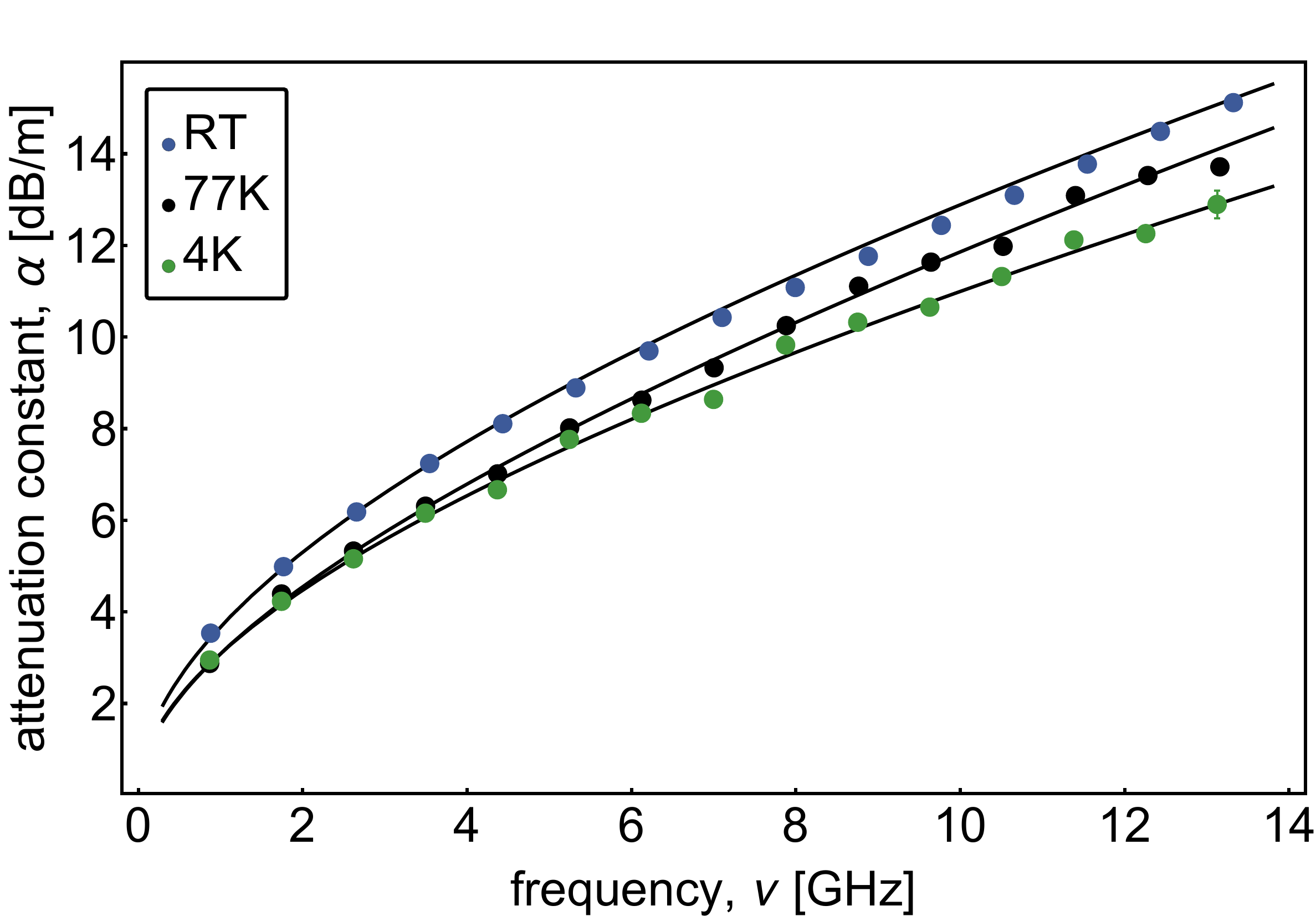}
\caption{Measured attenuation constant $\alpha$ (dots) versus frequency $\nu$ for a CC085SS at RT, approximately $77\,\rm{K}$ (LN2),  $4\,\rm{K}$ (LHe). The black lines are calculated from fits to the measured $Q_{\rm{i}}$ using the model of Eqn.~\ref{eqn:Qint2} as described in the main text.}
\label{fig:plotAlphaSSSS}
\end{figure}

We evaluate the loss of a $2.2\,\rm{mm}$ ($0.085\,\rm{in}$) diameter stainless steel outer and center conductor coaxial cable (CC085SS) typically used in cryogenic applications where loss is of little concern, i.e.~in input drive lines. At RT, approximately $77\,\rm{K}$ (LN2) and $4.2\,\rm{K}$ (LHe), the cable manufactured by Micro-Coax, Inc.~\cite{MicroCoax2016} shows the expected frequency dependent attenuation $\propto \sqrt{\nu}$ of a normal conductor decreasing with temperature (Fig.~\ref{fig:plotAlphaSSSS}). For example, we extract attenuation constants at $6\,\rm{GHz}$ of $9.7\,\rm{dBm/m}$ (RT), $8.7\,\rm{dBm/m}$ (LN2) and $8.2\,\rm{dBm/m}$ (LHe).

\begin{figure}[t]
\centering
\includegraphics[width=0.46\textwidth]{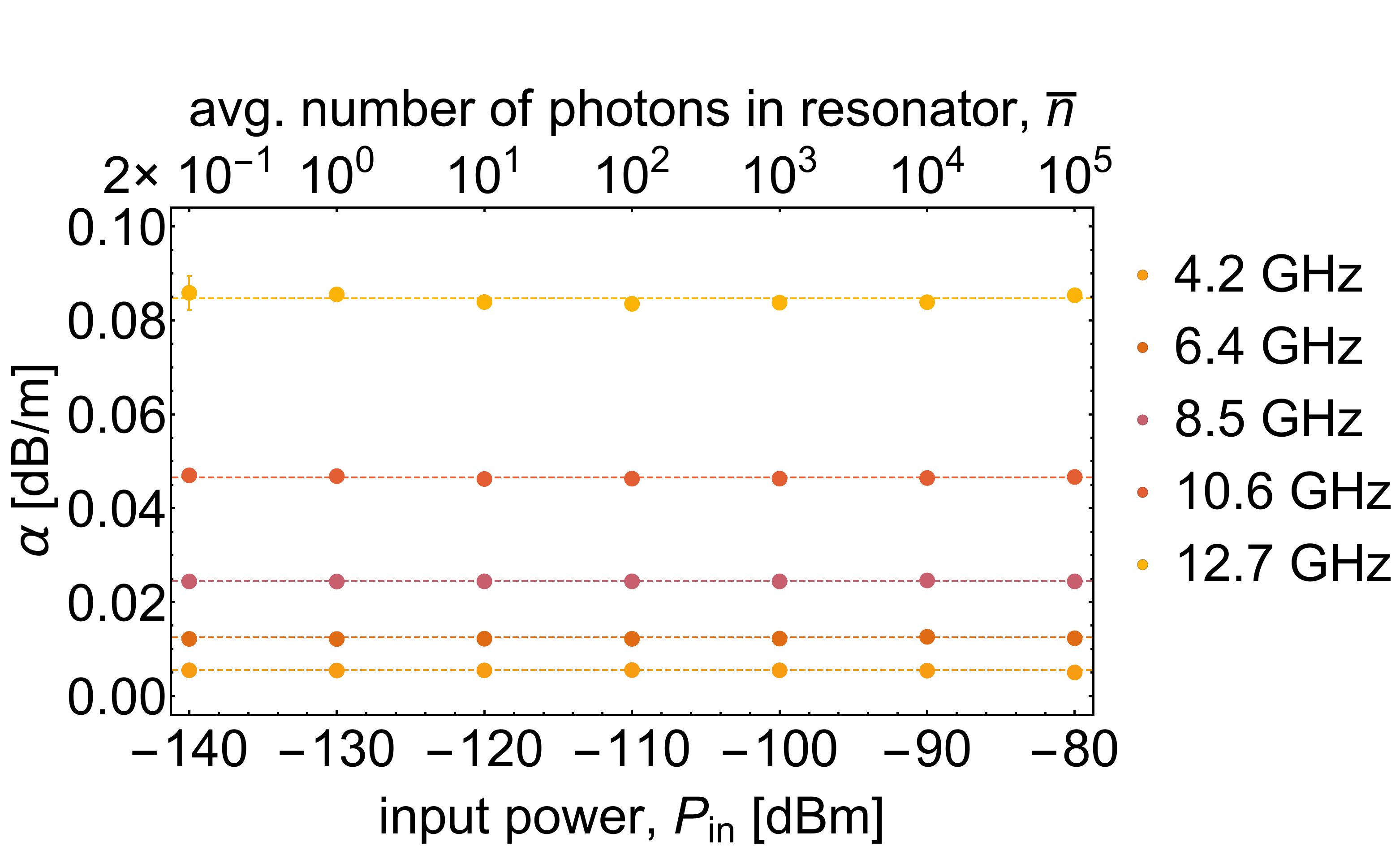}
\caption{Attenuation constant $\alpha$ (dots) versus input power at the waveguide $P_{\rm{in}}$ measured for frequencies between $4.2\,\rm{GHz}$ and $12.7\,\rm{GHz}$. The top axis indicates the average number of photons on resonance $\overline{n}$.}
\label{fig:PowerDepAlpha}
\end{figure}

\section{Power dependence of the attenuation constant}
\label{app:PowerDep}

We find the measured attenuation constant of CC085NbTi to be independent of the input power in a range from $-140$~to~$-80\, \rm{dBm}$ corresponding to an average photon number on resonance inside the resonator of $10^5$ to $0.2$ and an average resonator voltage $\left\langle V^2\right\rangle^{1/2}$ of approximately $10^{-6}$ to $10^{-4}~\rm{V}$ (Fig.~\ref{fig:PowerDepAlpha}). Our voltage range is comparable to that of Ref.~\cite{Martinis2005} in which a clear power dependence for Si$\rm{O}_2$ and Si$\rm{N}_x$ dielectric materials is observed and explained by the loss due to the coupling of microscopic two level systems (TLS) to the electromagnetic field within the resonator~\cite{Goetz2016} 
\begin{equation}
\delta_{\rm{TLS}}(P_{\rm{r}})=\frac{\delta_{\rm{TLS}}^0}{\sqrt{1+P_{\rm{r}}/P_{\rm{c}}}}
\label{eqn:LossTLS}
\end{equation}
with the low power TLS loss $\delta_{\rm{TLS}}^0$ and a characteristic power $P_{\rm{c}}$ depending on the dielectric material. Following this model we conclude that the field inside the dielectric material of the coaxial cable (ldPTFE) does not saturate the individual TLS in the measured power range~\cite{Von1977,Lindstrom2009}, since no power dependence of the attenuation of the coaxial cable is observed.

\section{Dependence on ambient magnetic field.}
\label{app:MagFieldDep}

\begin{figure}[h]
\centering
\includegraphics[width=0.48\textwidth]{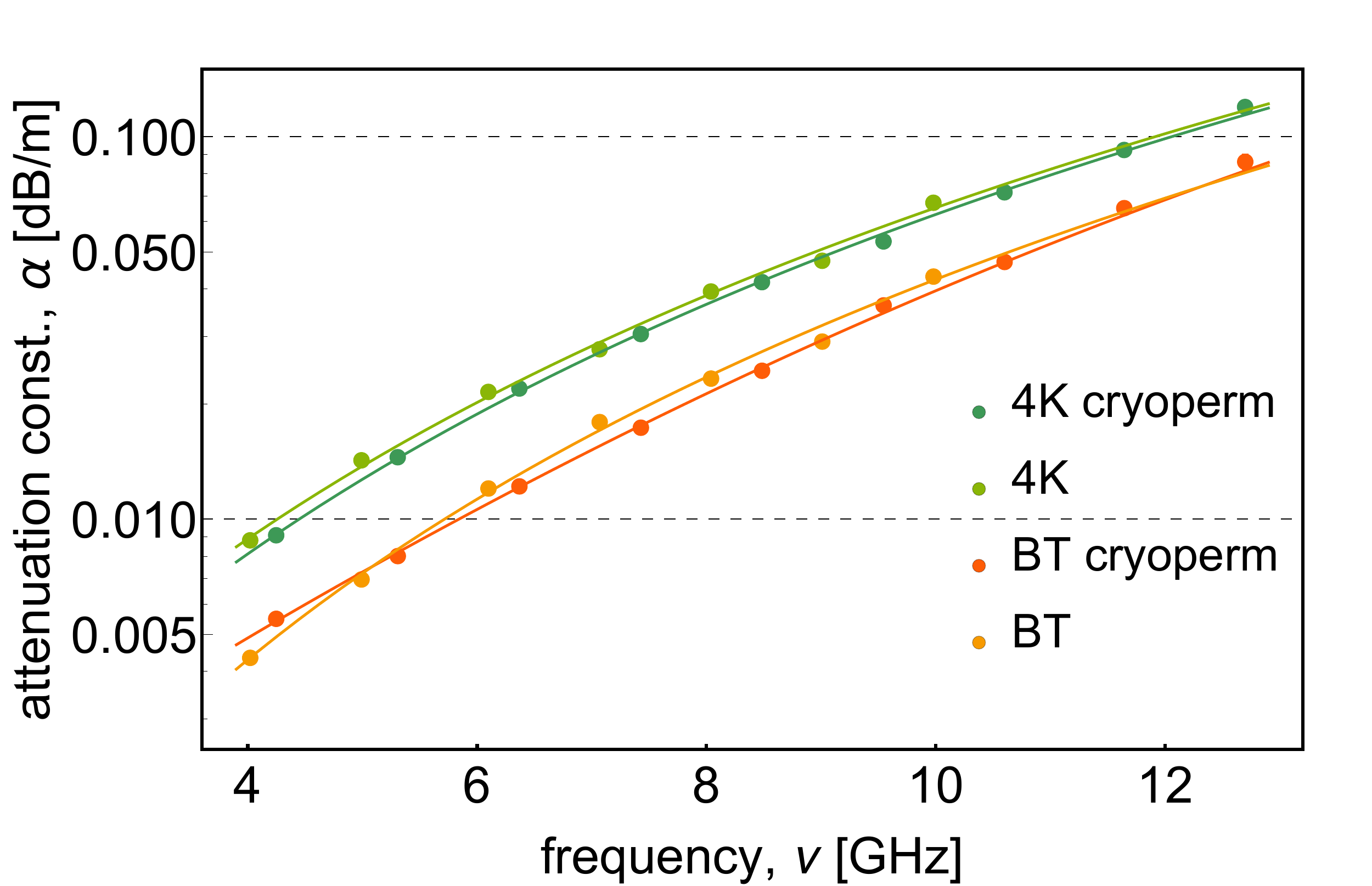}
\caption{Comparison of the attenuation constants measured at 4K and BT with and without cryoperm magnetic shielding for CC085NbTi. The solid lines are extracted from fits to the measured quality factors according to the model of Eqn.~\ref{eqn:Qint2}.}
\label{fig:plotAlphaCryoperm}
\end{figure}

We compare the extracted attenuation constants of CC085NbTi cables within (length 110 mm) and without (length 810 mm) a cryoperm magnetic shield and find no significant effect at 4K and BT (Fig.~\ref{fig:plotAlphaCryoperm}). Since we expect the internal loss to be the sum of the individual loss contributions we argue that the measured attenuation constants are not limited by an ambient magnetic field which is believed to be dominated by the isolators installed at the BT stage of the used dilution refrigerator system.

~\newline

\bibliographystyle{apsrev4-1}
\bibliography{attWaveguideRefDB}
\end{document}